\documentclass[12pt,preprint]{aastex}

\usepackage{emulateapj5}           % emulates original format

\slugcomment{Draft version
  from \today, accepted for ApJL}

\shorttitle{Outflow driving sources in Orion-KL}
\shortauthors{Beuther \& Nissen}

\begin{document}

\title{Identifying the outflow driving sources in Orion-KL}

%% Use \author, \affil, and the \and command to format
%% author and affiliation information.
%% Note that \email has replaced the old \authoremail command
%% from AASTeX v4.0. You can use \email to mark an email address
%% anywhere in the paper, not just in the front matter.
%% As in the title, you can use \\ to force line breaks.

\author{H.~Beuther$^1$ \& H.D. Nissen$^2$}
%\affil{Max Planck Institute for Astronomy, K\"onigstuhl 17, 69117 Heidelberg, Germany}
\altaffiltext{1}{Max Planck Institute for Astronomy, K\"onigstuhl 17, 69117 Heidelberg, Germany}
\altaffiltext{2}{Department of Physics and Astronomy, University of Aarhus, 8000 Aarhus C, Denmark}
\email{beuther@mpia.de, hdn@phys.au.dk}

\begin{abstract}
  The enigmatic outflows of the Orion-KL region have raised
  discussions about their potential driving sources for several
  decades. Here, we present C$^{18}$O(2--1) observations combined from
  the Submillimeter Array and the IRAM\,30\,m telescope. The molecular
  gas is associated on large scales with the famous
  northwest-southeast high-velocity outflow whereas the high-velocity
  gas on small spatial scales traces back to the recently identified
  submm source SMA1. Therefore, we infer that SMA1 may host the
  driving source of this outflow. Based on the previously published
  thermal and maser SiO data, source {\it I} is the prime candidate to
  drive the northeast-southwest low-velocity outflow. The source SMA1
  is peculiar because it is only detected in several submm wavelength
  bands but neither in the infrared nor cm regime. We discuss that it
  may be a very young intermediate- to high-mass protostar. The
  estimated outflow masses are high whereas the dynamical time-scale
  of the outflow is short of the order $10^3$\,yrs.
\end{abstract}


\section{Introduction}

The explosive large-scale northwest-southeast outflow in the Orion-KL
region is one of the best studied but probably also least understood
outflows in the sky. No other region shows that many H$_2$ bow shocks
distributed in a fan-like fashion with red- and blue-shifted emission
features on both sides of the central region (e.g.,
\citealt{schultz1999,nissen2007}). Furthermore, the molecular emission
as traced in CO shows extremely broad line-wings up to $>
50$\,km\,s$^{-1}$, but the morphological structure has so far been
unclear and difficult to match closely to the H$_2$ outflow (e.g.,
\citealt{chernin1996,rodriguez1999}). In addition to this outflow, also
known as the high-velocity outflow, there exists a second outflow in
the northeast-southwest direction which was first detected by H$_2$O
maser emission and which is often labeled as the low-velocity outflow
(e.g., \citealt{genzel1981,genzel1989}).

\begin{figure}
\includegraphics[angle=-90,width=0.48\textwidth]{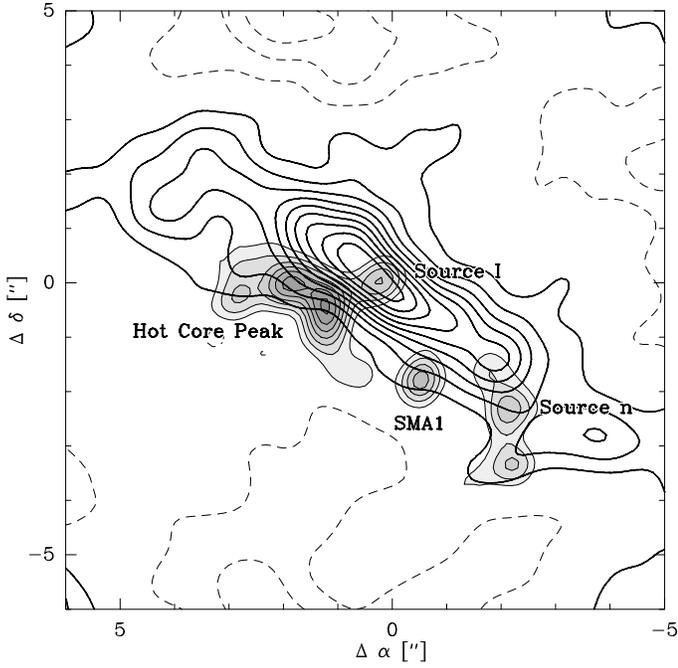}\\
\caption{SiO(8--7) integrated emission as contour overlay on the
  grey-scale 865\,$\mu$m submm continuum emission obtained with the
  SMA. This figure is an adaption of the 865\,$\mu$m continuum and
  SiO(8--7) line data from \citet{beuther2004g,beuther2005a}. The
  grey-scale continuum contours start at the $3\sigma$ level of
  0.105\,mJy\,beam$^{-1}$ and continues in $2\sigma$ steps. The SiO
  emission is contoured from $\pm$10 to $\pm$90\% (step $\pm$10\%) of
  the peak emission of 2.1\,Jy\,beam$^{-1}$ with full and dashed
  contours as positive and negative features (due to missing short
  spacings), respectively.}
\label{sio}
\end{figure}

The major question is which sources drive the different outflows.
While originally the infrared source IRc2 was a suspect, it was soon
realized that one of the prime powering sources of the region is the
very close-by radio source {\it I} \citep{menten1995}. This source is
the site of vibrationally excited SiO maser emission, and the
orientation of these masers allowed different interpretations of the
associated outflow direction. While early work favored source {\it I}
to drive the large-scale northwest-southeast outflow (e.g.,
\citealt{wright1995,greenhill1998}), more recent work on the maser
data indicates that it rather belongs to the low-velocity
northeast-southwest outflow \citep{greenhill2003}. The latter
interpretation finds additional support in thermal SiO observations
that show an elongated structure centered on source {\it I} with an
orientation also in the northeast-southwest
(\citealt{blake1996,beuther2005a}, see also Fig.~\ref{sio}). Since
thermal SiO emission is usually attributed to shocks within outflows
(e.g., \citealt{schilke1997a}), the combination of the thermal and
maser SiO emission strongly favors source {\it I} as the driving
source of the low-velocity northeast-southwest outflow. Source {\it n}
has also been discussed as a possible driving source of the
northeast-southwest low-velocity outflow. It shows an elongation along
position angle 100--130$^{\circ}$ in the mid-infrared which could be
due to a disk \citep{shuping2004,greenhill2004}, and a roughly
perpendicular elongation in the radio at 8.4\,GHz which may be caused
by an outflow \citep{menten1995}. The relationship between sources
{\it I} and {\it n} in driving the northeast-southwest outflow(s) is
unknown (for a discussion see, e.g., \citealt{shuping2004,nissen2007}).

That leaves as an open question which source is the culprit for
driving the large-scale high-velocity northwest-southeast outflow?
\citet{bally2008} recently suggested that this outflow may be powered
by the dynamical decay of a massive star system, but other scenarios
are possible as well. Based on low-resolution data,
\citet{devicente2002} proposed the existence of another luminous
source approximately $2''$ south of source {\it I}.  This source was
unambiguously identified in the first sub-arcsecond resolution
865\,$\mu$m submm continuum image obtained with the Submillimeter
Array (SMA), and henceforth labeled SMA1 (\citealt{beuther2004g}, see
also Fig.~\ref{sio}).  Follow-up SMA observations at 440\,$\mu$m
detected the source at these shorter wavelengths as well
\citep{beuther2006a}. In this work we show that the high-velocity gas,
as traced by C$^{18}$O, apparently traces back to this relatively
unknown source rather than to sources {\it I} or {\it n}.

\section{Observations}
\label{obs}

Orion-KL was observed with the SMA in the compact configuration
simultaneously at 690\,GHz and 230\,GHz on 2005 February 19th. To
accomplish the high-frequency observations the weather conditions were
excellent with zenith opacities $\tau(230\,\rm{GHz})$ between 0.03 and
0.04 throughout the night. The phase center was the nominal position
of source {\it I} R.A.(J2000) 05h35m14.50s and decl.(J2000)
05$^{\circ}22'30.''45$, and the $v_{\rm{lsr}}$ was
$\sim$5\,km\,s$^{-1}$. The double-sideband receivers covered the
frequency ranges 218.85 to 220.85\,GHz and 228.85 to 230.85\,GHz. The
phase was calibrated with regularly interleaved observations of the
quasar 0607-157, flux and bandpass calibration were performed with
Callisto observations.  The high-frequency data along with more
observational details are published in \citet{beuther2006a}, here we
are only interested in a small subset of the low-frequency data.
Complimentary C$^{18}$O(2--1) data were obtained with the 9 pixel HERA
array on the IRAM 30\,m telescope in September 2007. The observations
covered a region of $1'\times 1'$ around source {\it I} and were
conducted in the on-the-fly mode with on-source integration times per
position of $\sim$3.7\,secs. The OFF-source reference position was
$15'$ east of source {\it I}. The single-dish data were converted to
visibilities and subsequently processed with the SMA data within the
MIRIAD package using standard tools (e.g., UVMODEL). The weighting of
the two datasets was chosen to recover large-scale emission but
maintain at the same time the high spatial resolution to resolve
small-scale structure.  The synthesized beam of the combined data is
$4.2''\times 3.3''$, and the $1\sigma$ rms per 5\,km\,s$^{-1}$ channel
is $\sim$50\,mJy\,beam$^{-1}$.

\section{Results}
\label{results}

\begin{figure*}
\includegraphics[angle=-90,width=0.98\textwidth]{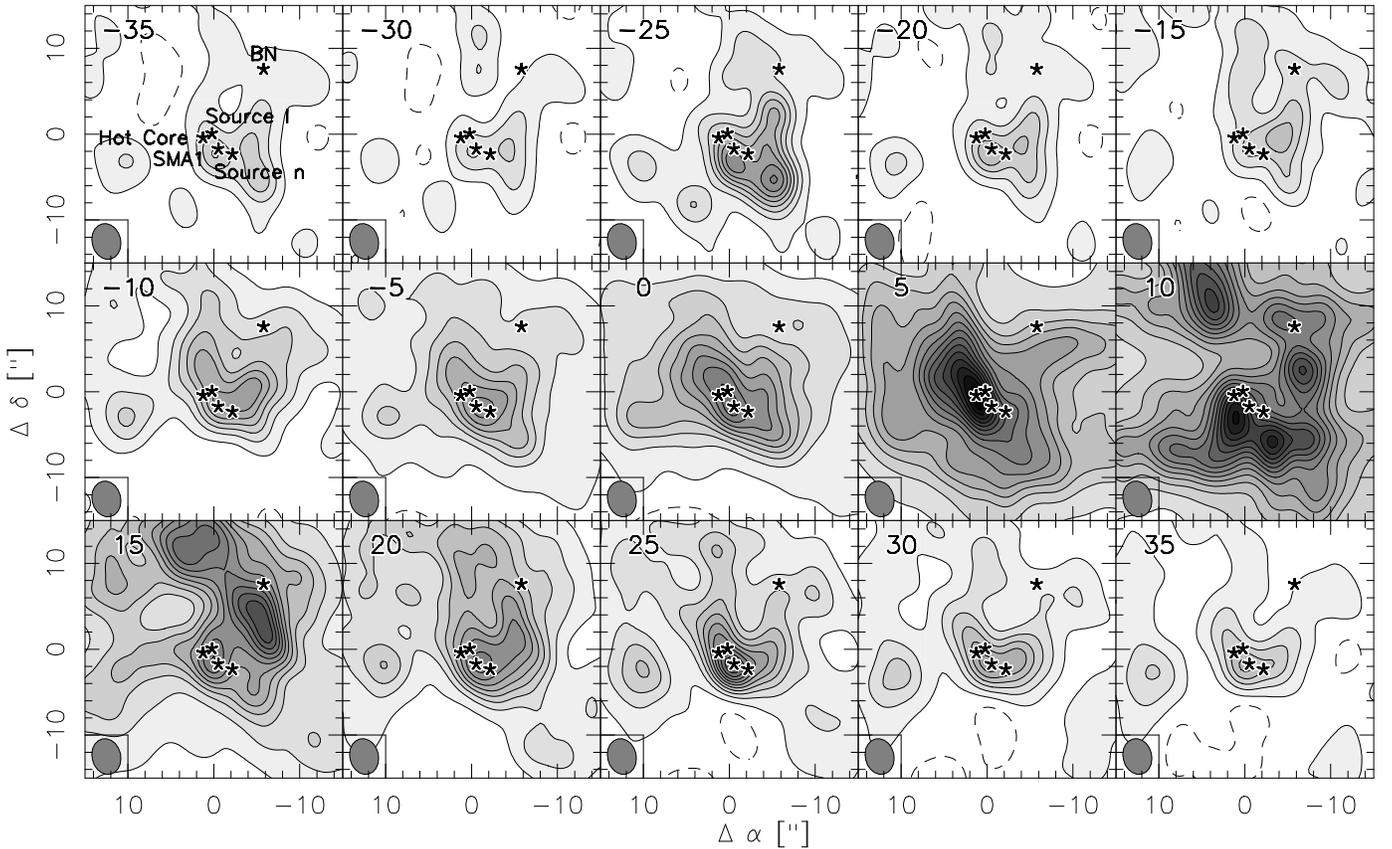}\\
\caption{C$^{18}$O(2--1) channel map with 5\,km\,s$^{-1}$ of the
  combined SMA+30m data. The contours start at the $3\sigma$ levels of
  150\,mJy\,beam$^{-1}$ and continue in $6\sigma$ steps for all panels
  except of the 0, 5 and 10\,km\,s$^{-1}$ panels that follow 12, 12
  and $14\sigma$ steps for clarity reasons. Negative features are draw
  as dashed contours. The stars mark the positions of the five main
  continuum source with labels in the top-left panel. The synthesized
  beam is shown at the bottom-left of each panel.}
\label{channel}
\end{figure*}

The whole low-frequency SMA dataset will be presented in a forthcoming
publication by Nissen et al.~(in prep.), here we are only interested
in the C$^{18}$O(2--1) emission. Although the SMA and IRAM\,30\,m
observations also covered the more abundant isotopologues
$^{12}$CO(2--1) and $^{13}$CO(2--1), in Orion-KL it turns out that
because of their high optical depth these two lines strongly suffer
from confusion problems with the ambient cloud complicating any
interpretation of the data. In contrast to that, the rarer
isotopologue C$^{18}$O(2--1) penetrates the ambient cloud more deeply
and hence traces the outflow associated gas far better, allowing us a
deeper analysis of the outflow structure, its mass flow rate,
kinematics and energetics, as well as pinpointing the likely driving
source.  Fig.~\ref{channel} presents the combined SMA+30\,m
C$^{18}$O(2--1) channel map. Excluding the channels with strongest
emission closest to the $v_{\rm{lsr}}$ (approximately four channels
between 0 and 15\,km\,s$^{-1}$), two main features can be
distinguished in the remaining channels:

(1) The large-scale gas emission exhibits a V-shaped cone-like open
structure to the northwest, particularly prominent in the red-shifted
channels $\geq 20$\,km\,s$^{-1}$ but also visible in the blue-shifted
channels between $-20$ and $-10$\,km\,s$^{-1}$. This cone-like
structure approximately encompasses the main features of shocked
strong H$_2$ emission in this direction (Fig.~\ref{c18o_h2}). Finding
both blue- and red-shifted gas to the northwest indicates a wide
opening angle of the flow where the main outflow axis is not too far
from the plane of the sky. The less pronounced emission to the
southeast was already previously observed, and \citet{chernin1996}
argue that the outflow may be blocked in this direction by the hot
core\footnote{The hot core traced in NH$_3$ \citep{wilson2000}
  actually extends further southeast of SMA1 than indicated by the hot
  core continuum peak in Fig.~\ref{sio}. The proposed blocking of the
  outflow by the hot core may also be responsible for some of the
  large-scale asymmetries of the H$_2$ emission.}. It is interesting to
note that the main southeastern emission features are well correlated
with a lack of H$_2$ emission indicating particularly high column
densities in this direction. Furthermore, we find high-velocity gas
toward the southwest and northeast (Fig.~\ref{c18o_h2}). Since
\citet{nissen2007} find H$_2$ emission with velocities up to
40\,km\,s$^{-1}$ toward the southwest as well, it is likely that this
high-velocity gas is associated with the low-velocity outflow.

(2) We find that the high-velocity gas on the blue- and red-shifted
side of the spectrum traces back to the source SMA1. Although the
synthesized beam is larger than the spatial separation of source {\it
  I} and SMA1 ($\sim 2''$or $\sim$830\,AU), the peak positions can be
determined with a much higher spatial accuracy depending on the
signal-to-noise ratio S/N. For point sources, one can in principle
achieve an accuracy of 0.45\,HPBW/(S/N) \citep{reid1988}, however,
this may be less applicable for extended emission like C$^{18}$O in
this case. The channel with the lowest S/N ratio still detects the
main peak position toward SMA1 at higher than $15\sigma$ values (at
$-30$\,km\,s$^{-1}$). For a point source, this would imply that the
peak positions could be determined to better than $0.125''$. Although
the C$^{18}$O emission is more extended, the shape of the
high-velocity gas close to SMA1 can to first order be described by a
two-dimensional Gaussian. Therefore, being more conservative for the
C$^{18}$O high-velocity gas, it should still be possible to infer the
emission peak positions to an accuracy of better than $1''$. Based on
this, the blue- and red-shifted high-velocity gas is spatially mainly
associated with SMA1, and not with the previously often discussed
sources {\it I} or {\it n}.

The blue- and red-shifted maps shown in Figure \ref{c18o_h2}
(integrated from $-35$ to $-20$ and from 22.5 to 37.5\,km\,s$^{-1}$,
respectively) can be used to calculate the masses, kinematic and
energetic parameters of the outflow following \citet{beuther2002b}
(time-scale $t$, outflow rate $\dot{M}_{\rm{out}}$, momentum rate $p$,
and energy $E$). Assuming optically thin C$^{18}$O(2--1) emission at
30\,K with a C$^{18}$O/H$_2$ conversion factor of $1.7\times 10^{-7}$
\citep{frerking1982}, the derived values are given in Table
\ref{parameters}.  While the masses have uncertainties of
approximately a factor 5, the kinematic and energetic values are
uncertain within an order of magnitude \citep{cabrit1990}. The derived
masses, outflow and momentum rates are at the upper end of typically
found massive outflows, whereas the time-scale is relatively short
(compared with the outflow sample in \citealt{beuther2002b}).

\section{Discussion and Conclusions}

The large-scale cone-like morphology of the C$^{18}$O outflow
approximately following the H$_2$ bow-shock emission in the northwest
direction suggests that C$^{18}$O and H$_2$ are tracing the same
outflow. Furthermore, the spatial association of the C$^{18}$O
high-velocity emission with the submm source SMA1 indicates that SMA1
may harbor the driving source of the enigmatic large-scale Orion-KL
outflow.  Only the combination of the single-dish 30\,m data, tracing
the large-scale structure, with the SMA high-spatial-resolution
observations allows us to see all relevant structures and to trace the
outflow back to pinpoint its origin.

In combination with the thermal and maser SiO data, we suggest that
the two driving sources of the two known outflows can be identified:
(1) SMA1 may host the driving source of the large-scale
northwest-southeast high-velocity outflow, whereas (2) source {\it I}
is likely the origin of the northeast-southwest low-velocity outflow.

The nature of source {\it I} has been subject to many investigations
for more than a decade (e.g.,
\citealt{menten1995,beuther2006a,reid2007}), however, we do not know
much about SMA1. So far SMA1 has only been detected at submm
wavelengths (865 and 440\,$\mu$m, \citealt{beuther2004g,beuther2006a})
but neither at shorter infrared nor at longer cm wavelengths (e.g.,
\citealt{menten1995,greenhill2004}). The submm-only detection is
indicative of a very young age of the source which is supported by the
short dynamical time-scale of the molecular outflow. Although deriving
a luminosity with just two data-points is impossible, based on the
large outflow masses, it is likely that SMA1 is in the process of
forming a massive star. The non-detection at cm wavelengths could
either be due to quenching of an underlying hypercompact H{\sc ii}
region by very high accretion rates and/or gravitational trapping of
the ionized gas (e.g., \citealt{walmsley1995,keto2003}), or the source
is still in such a young evolutionary phase that the central object
has not yet enough ionizing flux to produce a detectable hypercompact
H{\sc ii} region. In the framework of the recently proposed
evolutionary sequence by \citet{beuther2006b}, it may be an
intermediate-mass protostar destined to become massive in the future.

Assuming optically thin dust emission at 100\,K with a dust opacity
index $\beta =2$ (e.g., \citealt{hildebrand1983,beuther2002a}), the
spatially resolved submm fluxes \citep{beuther2004g,beuther2006a}
correspond to a gas mass and column density of $\sim$0.1\,M$_{\odot}$
and $3\times 10^{24}$, respectively. While the large column densities
confirm that this source is undetectable at near-infrared wavelengths
with current instrumentation, at first sight the mass appears small.
However, as discussed in \citep{beuther2004g}, more than 90\% of the
total flux is filtered out by the SMA observations, and hence the
mass estimates are correspondingly unreliable. The total luminosity
of the region of about $10^5$\,L$_{\odot}$ (e.g.,
\citealt{menten1995}) cannot be produced by the sources previously
discussed in the literature (e.g., \citealt{greenhill2004}), and it is
well possible that SMA1 contributes a significant fraction of the
overall Orion-KL luminosity.

We note that the short dynamical outflow time-scale of the order
$10^3$\,yrs is consistent with both recently discussed scenarios for
the proper motions in this region: \citet{tan2005} proposed that
the recent passage of the BN-object has triggered the outflow burst in
this region, whereas \citet{rodriguez2005} and \citet{gomez2005}
suggest that a dynamically decaying multiple system is at the root of
the observed proper motions. Both events are suggested to have taken
place about 500\,yrs ago. Our data do not allow to differentiate
between these scenarios, however, it is interesting that all recent
observations are indicative of strong activity within the last
$10^3$\,yrs.

\acknowledgments{H.B. acknowledges financial support by the
  Emmy-Noether-Programm of the Deutsche Forschungsgemeinschaft (DFG,
  grant BE2578). H.D.N. acknowledges the support of the Aarhus Centre
  for Atomic Physics (ACAP), funded by the Danish Basic Research
  Foundation and the financial support from the Instrument Centre for
  Danish Astrophysics (IDA), funded by the Danish National Science
  Committee (FNU). }

%%%%%%%%%% to create the bibliography
%\bibliography{/home/beuther/tex/bibliography}
%\bibliography{/Users/henrikbeuther/paper/bibliography}
%\bibliographystyle{aa}    % this does the style, aa.bst necessary

\clearpage

\begin{table}[ht]
\caption{Outflow parameter}
\begin{tabular}{lr}
\hline 
$M_{\rm{blue}}$ [M$_{\odot}$] & 33 \\
$M_{\rm{red}}$ [M$_{\odot}$] & 81 \\
$M_{\rm{total}}$ [M$_{\odot}$] & 124 \\
$t$ [yr] & 971 \\
$\dot{M}_{\rm{out}}$ [M$_{\odot}$\,yr$^{-1}$] & $1.2\times 10^{-1}$ \\
$p$ [M$_{\odot}$\,km\,s$^{-1}$] & 3980 \\
$E$ [erg] & $1.4\times 10^{48}$ \\
\hline 
\end{tabular}
\label{parameters}
\end{table}

\clearpage

\begin{figure*}
\caption{Overlay of the H$_2$ emission in grey-scale
  \citep{nissen2007} with the blue- and red-shifted C$^{18}$O(2--1)
  emission in color-contours. The integration ranges are
  [$-$35.0/$-$20.0] and [22.5,37.5]\,km\,s$^{-1}$, respectively. The
  contours are in $3\sigma$ steps of 150\,mJy\,beam$^{-1}$. The stars
  mark the positions of the same continuum sources as in Figure
  \ref{channel}, the synthesized beam is shown at the top-right of
  each panel.  The large circle in the left panel shows the size of
  the SMA primary beam, whereas the squared box marks the size of the
  inlay-box shown in the right panel. The outflow directions of the
  high- and low-velocity outflow are sketched in the left panel.}
\label{c18o_h2}
\end{figure*}

\end{document}